# SNS Timing system

J R. Nelson LANL, B. Oerter BNL, T Shea ORNL, C. Sibley ORNL

Abstract

A modern physics facility must synchronize the operation of equipment over a wide area. The primary purpose of the site wide SNS synchronization and timing system is to synchronize the operation of the LINAC, accumulator ring and neutron choppers and to distribute appropriate timing signals to accelerator systems, including the Injector, LINAC, Accumulator Ring and Experimental Facilities. Signals to be distributed include the ring RF clock, real-time timing triggers, machine mode and other informational events. Timing triggers and clocks from the SNS synchronization and timing system are used to synchronize hardware operations including the MEBT beam chopper, RF turn on, synchronous equipment state changes, as well as data acquisition for power supplies and beam diagnostics equipment. This paper will describe the timing equipment being designed for the SNS facility and discuss the tradeoffs between conflicting demands of the accelerator and neutron chopper performance due to AC power grid frequency fluctuations.

## 1 REQUIRMENTS

### 1.1 Synchronization of Neutron Choppers[1]

The neutron choppers are very high inertia mechanical rotors with collimators that chop the neutron beam. To prevent modulation of the neutron energy spectrum, the chopper operation must be synchronous with the arrival of protons on the target. Thus the phase of the chopper is fixed with respect to extraction while the phase of a chopper is independent of the operation of the LINAC. For the several types of choppers, Fermi, $T_0$, and bandwidth limiting choppers, the most stringent desired timing accuracy for the synchronization is about ±0.5µs. The required accuracy is still to be determined, but is in the ±1 µs range.

### 1.2 Synchronization of LINAC

In order to inject approximately 1060 turns into the ring, the beam injection into the LINAC must begin about 896 µs before extraction. For a super conducting LINAC, the klystron modulators must be pulsed about 400 µs earlier than the beam. The beam diagnostics group have requested a pulse, approximately 5500 turns before the ring extraction time.

Klystons in the LINAC require synchronization with the power grid. Failure to maintain this synchronization results in non-linear performance and beam losses. The beam in the LINAC injector is chopped by a gated electrostatic focusing electrode in the LEBT and a traveling-wave deflector (beam chopper) in the MEBT. Each of these systems, and especially the beam chopper in the MEBT, must be synchronized to the ring period. The reason for chopping the beam is to produce a gap in the beam current that is synchronized to the ring period.

The beam occupies about 67% of the ring circumference, and the remainder of the ring (the beam gap) is free of beam (to about 1 part in $10^4$) in order to accommodate "clean" extraction. The extraction from the ring must be synchronized to the beam to about ± 5 ns (the beam gap is about 300 ns long).

### 1.3 Ring Revolution period

During normal operation, tuning the ring by changing the B field (which changes the orbit radius), or tuning the LINAC to change the injection energy, or a combination of both, can change the revolution period by up to ± 1 ns. Although 2 ns is not a large number, the cumulative timing error over a normal accumulation cycle of 1060 turns exceeds 2 µs, more than one complete ring period. This is much larger than the desired synchronization accuracy of ± 5 ns for accumulating 1060 turns of beam in the ring. A beam synchronous timing system tied to the rotation frequency of the ring will compensate for this variation in fill time and maintain a ±5 ns timing accuracy. The system designed for SNS will operate over an energy range of 950 MeV to 1.3 GeV without modification

## 2 BEAM DIAGNOSTICS

### 2.1 Ring

The beam diagnostics in the ring measure the revolution of the beam during the accumulation cycle. The ring beam diagnostic measurements include global systems such as a beam closed-orbit measurement (BPM) system, and local systems such as the azimuthal distribution of beam charge density (current monitor). Global and local refer respectively to systems that are widely distributed around the ring, or are needed in only one or two locations. These systems need to be synchronized to the beam in the ring to within about ±5 ns, even though the accumulation period may vary by 2000 ns or more. Diagnostics in the beam line between the LINAC and ring requires synchronization to the ring period. Diagnostics in the beam line between the ring and the spallation target need only a single pretrigger a few turns before extraction.

## 2.2 LINAC

Although the LINAC RF is not synchronized to the ring, the periodic beam gap in the LINAC is. Because of the effect of the beam gap on the performance of the LINAC beam diagnostics, the sampling of the beam diagnostics signals from all the diagnostics needs to be synchronized relative to the beam gap. In particular, beam current, beam position, and beam synchronous phase measurements would benefit from being synchronized relative to the beam gap. Thus the LINAC diagnostics also need to be synchronized to the beam in the ring. A 60-ns timing granularity (not jitter) of a timing signal relative to the beam gap is adequate (the beam gap is 300 ns long, and the beam mini pulse between beam gaps is 645 ns long).

# 3 Timing System Architecture

## 3.1 Synchronization

Experience at other spallation sources has shown that the LINAC, ring, and neutron choppers can synchronize operation with each other and follow trends in the phase fluctuations of the ac power grid. Each subsystem operates as a slave to a master timing generator which smoothes phase fluctuations present in the power grid. At SNS the competing demands of the LINAC that limits phase differences with respect to the power grid and of the neutron choppers that limit the acceleration or deceleration of high-inertia rotors can be simultaneously satisfied within a "phase" window of ±500 µs measured with respect to the power grid.

## 3.2 PLL

The master timing generator will implement a phase-locked loop (PLL) that will follow the phase fluctuations of the power grid and produce the Cycle Start signal with "smoothed" phase fluctuations. The residual fluctuations should not exceed the ability of the neutron choppers to maintain phase lock with Cycle Start or with signals derived from the timing distribution system. Similarly the residual phase fluctuations must not exceed ±500 µs to ensure correct operation of the klystrons in the LINAC.

The ac-line synchronization requirement should be established by an operational parameter varied under computer control. This parameter characterizes the variance in the time interval between the zero crossing and the Cycle Start signal, even during line frequency transients. Over its full range this parameter will provide for phase coupling from that is essentially identical to the power grid through phase coupling with a standard deviation of 125 µs. While this parameter may be adjusted for certain measurements, normally its value would be fixed.

To synchronize the accelerator systems and the neutron choppers, a phase-stable timing signal is generated and distributed around the facility. **Cycle Start** will be defined. **Cycle Start** shall be generated some number, of ring-rf cycles before extraction. The actual number might vary depending on the number of micropulses injected into the ring. The phase of the ring RF will be resynchronized at **Cycle Start** while the accelerator is idle and the ring is empty. Once established at **Cycle Start**, the phase and frequency are held fixed until after the beam is extracted.

Should the master timing generator lose synchronization as indicated by the **Cycle Start** signal drifting too far from the zero crossing, then a sync lost pulse will be generated for each cycle until synchronization is regained. Conditions for defining lost synchronization are TBD.

To synchronize extraction of the beam from the storage ring with the rotation of neutron choppers, the master timing generator will first determine when beam should be extracted for the next beam pulse. Then the timing generator must schedule generation of the Cycle Start signal allowing for $N_{cs}$ ring-rf cycles prior to extraction. The Cycle Start signal will reset the phase of the ring rf to zero.

## 3.3 Event Link Timing Distribution

The most convenient way to *broadcast* timing signals to many hundreds of clients distributed over a kilometer or more of accelerator and beam lines is to encode all the timing signals using a self-clocking scheme in which the timing signals are encoded on the carrier, and distributed to all clients. If the carrier frequency on which the events are encoded, is synchronous with the beam in the ring, rather than a fixed frequency, then timing triggers and delays derived from the carrier frequency maintain the proper relationship with the beam revolution period. This eliminates the need to adjust timing delays due to changes in the ring revolution period. SNS will use a carrier frequency of 16 times the ring revolution period.

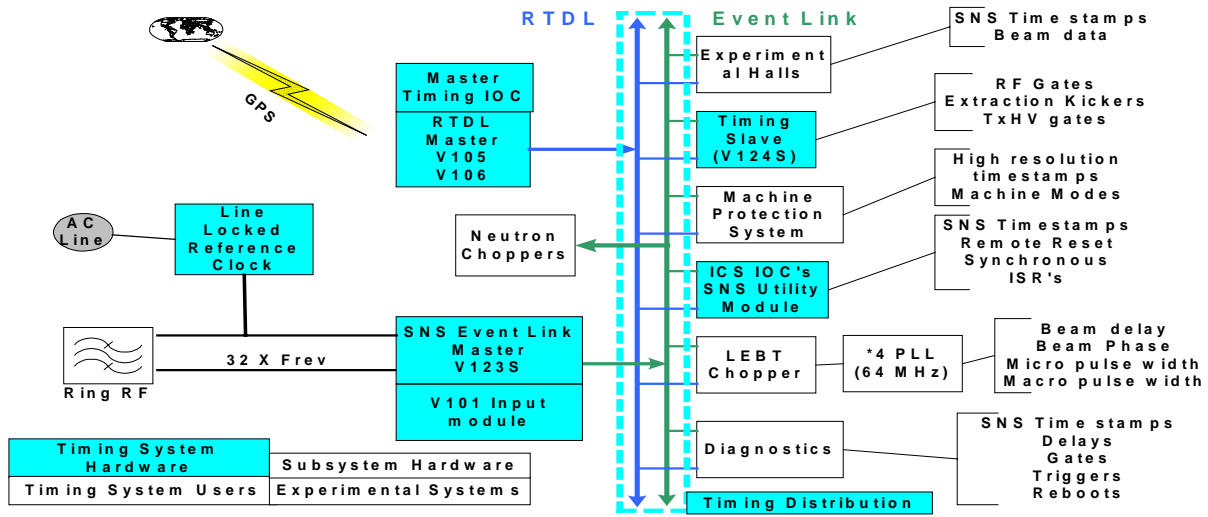

Figure 1 SNS Timing System Distribution

## 3.3 RTDL System

The RTDL system is a master slave system used to broadcast data in real time. IOCs will have access to RTDL broadcasts through the Utility Module. A maximum of 256 data frames can be defined. The RTDL encoder contains a list of frames to be transmitted. All frames contained in the list are sent out prior to the start of each accumulation cycle. A list of the defined frames is shown below.

1. Time of day (IOC timestamps)
2. Ring revolution period (in ps.)
3. Operating Mode
4. 60 Hz phase difference
5. Beam Parameters for LEBT chopper
6. Previous beam pulse data
7. Data acquisition mode
8. Beam profile ID
9. Transmitter, RF Gates ON/OFF
10. IOC reset address

## 3.3 Time stamp

A commercial VME module resident in the RTDL chassis receives time of day information from a GPS receiver. At **RTDLStart** the time will be updated in the RTDL system for broadcast to all receivers. Thus each machine cycle shall have a unique time of day timestamp. For systems requiring a timestamp with greater resolution within a machine cycle, hardware counting at some multiple of the ring revolution frequency shall provide the additional fractional machine cycle component of the time stamp.

## 3.4 Utility Module

Each IOC will contain a Utility Module. The Utility Module provides a number of VME chassis services including receiver circuits for the Eventlink and RTDL links. The Utility Module can be configured to initiate VMEbus interrupts on the detection of specified events. Data transmitted on the RTDL are stored in local memory. One of the RTDL frames provides remote reset of VME chassis. All Utility Modules monitor this frame for their remote reset address. When the reset code matching an IOCs Utility Modules preprogrammed reset address is received, the Utility module asserts the VMEbusSYSRESET/ signal

## 4 Status - Conclusion

The SNS Timing System design is based on the RHIC timing system. Using this proven design minimizes new design risks and will result in a system with proven performance and reliability. Modifications for the SNS requirements have been straightforward. Prototypical systems are running at BNL and SNS. Integration Software (Run Permit System, see SNS Run Permit System Poster at this conference) in progress using hardware and software supplied by BNL.